\documentclass[show,showpacs,amsmath,amssymb,twocolumn,superscriptaddress]{revtex4}
\usepackage{graphicx,float}
\usepackage[all]{xy}
\usepackage{amsmath}
\usepackage{amssymb}
\usepackage{color}

\newcommand{\bes}{\begin{subequations}}
\newcommand{\ees}{\end{subequations}}
\def\ben{\begin{eqnarray}}
\def\een{\end{eqnarray}}
\def\be{\begin{equation}}
\def\ee{\end{equation}}
\def\sech{\text{sech}}
\def\sech{\textrm{sech}}

\def\LL{{\cal L}}
\begin{document}
\title{First Order Formalism for Thick Branes in Modified Teleparallel Gravity}

\author{Roberto Menezes}
\affiliation{Departamento de Ci\^encias Exatas, Universidade Federal da Para\'\i ba, 58297-000 Rio Tinto, PB, Brazil.}

\affiliation{Departamento de F\'\i sica, Universidade Federal de Campina Grande, 58109-970 Campina Grande, PB, Brazil}


\date{\today}
\begin{abstract}
 This work deal with braneworld scenarios with generalized teleparallel gravity. We extend a recent investigation, where the model studied was described by $F(T)=T+k T^n$.  In the current study, we introduce the first-order formalism to find analytical solutions for models that include scalar field with standard and generalized dynamics. In particular, we describe the interesting case in which the brane splits, due to the parameters that control deviation from the standard model. 
 \end{abstract}
\maketitle

\section{Introduction}
Teleparallel gravity is equivalent to the general gravity~\cite{Moller}. The tetrad field $h^M_a$ is used to construct the Weitzenb\"ock connection 
$\tilde \Gamma^{M}_{\,\,\,{NS}}=h^M_a \partial_N h^a_S$, which is distinct from Levi-Civita connection $\Gamma^{M}_{\,\,\,{NS}}$ \cite{Weit}. While the Levi-Civita connection has curvature and no torsion, the Weitzenb\"ock connection has torsion and no curvature. The tetrad field generates a teleparallel structure on spacetime, related with the gravitational field $g_{MN}=\eta_{ab}h^{a}_{\,\,\,M} h^{b}_{\,\,\,N}$, where the indices with up and lower cases represents the spacetime and tangent space,  respectively and $\eta_{\mu\nu}$ is the Minkowski space. The relation between  Weitzenb\"ock and Levi-Civita connections is given by $ K^{P}_{\,\,\,MN}=\tilde\Gamma^{P}_{\,\,\,{MN}}-\Gamma^{P}_{\,\,\,{MN}}$, with 
\ben
K^{P}_{\,\,\,MN}&=&\frac12 \left(T_{M\,\,\,N}^{\,\,\,P}+T_{N\,\,\,M}^{\,\,\,P}-T_{\,\,\,MN}^{P}\right),\\
T^{P}_{\,\,\,MN}&=&\tilde\Gamma^{P}_{\,\,\,MN}-\tilde\Gamma^{P}_{\,\,\,NM},
\een
where $K^{P}_{\,\,\,MN}$ and $T^{P}_{\,\,\,MN}$ are the contortion and torsion, respectively. 

The 5D action of the teleparallel gravity is \cite{Lagrangian}
\be 
{\cal S}_T= -\frac{1}{4} \int d^5x \, h\, T,
\ee
where $T$ scalar is defined as
\be
T=\frac14 T^{P}_{\,\,\,MN}  T_{P}^{\,\,\,MN}+\frac12 T^{P}_{\,\,\,MN}  T^{NM}_{\,\,\,\,\,\,\,\,P}-T_{PM}^{\,\,\,\,\,\,\,\,\,P} T^{NM}_{\,\,\,\,\,\,\,\,\,\,N}
\ee
and $h=\det(h^a_M)$. Here we take $4\pi G=c^4$. This action is identical to Einstein-Hilbert action for the general gravity
\be 
{\cal S}_R= \frac{1}{4} \int d^5x \, \sqrt{-g}\, R,
\ee
where $R$ is the Ricci scalar (see Refs.\cite{Lagrangian} for reviews). This equivalence is not preserved in scenarios where the gravity is modified. In general, these two actions
\ben
S_R=\frac{1}{4} \int d^5x \, \sqrt{-g} \,f(R),\,\,\,\,\,\, S_T=-\frac14\int d^5x \, h\, f(T) 
\een 
lead to two quite different equations for gravity. Recently, theories with $f(T)$ contributions have been considered in cosmological scenarios as an interesting way for cosmic accelerating expansion without dark energy~\cite{Frevision}. Formal aspects for this type of models were studied in several references~\cite{Others}.

In the recent work \cite{Yang:2012hu}, the authors investigated the presence of thick brane solution for the model with $F(T)=T + k T^n $. They obtained analytical solutions from a given warp factor. In special they have shown that the appearance of double-kink solutions leads to the splitting of the  brane. This occurs when the parameter $k$ is large.
 In the current work, we extend the investigation to more general models, including the presence of a scalar field $\phi$ with standard and generalized dynamics, as we explain in the next Section.

The braneworld consists in a domain wall embedded in the higher dimensional bulk. The defect represents the three-dimensional universe. For more than one decade, distinct braneworld scenarios have been studied. In this environment, relevant issues which can be discussed are the gauge hierarchy and the cosmological constant problems, for instance~\cite{Brane1, Brane2}. 

The aim of this paper is to investigate braneworld scenarios with $f(T)$ gravity. We use the first-order formalism (fake superpotential method) to find
analytical solutions for the coupled scalar-gravity equations. This established method has been used extensively in several situations~\cite{Brane2,Brane3,Brane4, Bazeia:2013uva}. Here we investigate applications for generalized models, with the scalar field having standard and modified dynamics, within the context of generalized teleparallel gravity.

\section{The Model}
We start with the 5D action
\be\label{themodel}
S=\int d^5x \,h\,\left[-\frac14 (T+ kT^{n})+{\cal L}(\phi,X)\right],
\ee
where $X=-\frac12 \nabla_A \phi \nabla^A \phi$. In this work, we introduce the first-order formalism for a generic Lagrangian density  of a scalar field ${\cal L}(\phi,X)$. As example, we focus our attention in two models
\bes
\ben\label{standardL}
\LL_1&=&X-V(\phi),\\
\LL_2&=&X+\alpha[(1+bX)^n-1]-V(\phi),\label{modifiedL}
\een
\ees
where $b>0$ is positive parameter and $\alpha$ is a real parameter. The first model (Model I) is the standard model that was studied in Ref.~\cite{Yang:2012hu}. The second model (Model II) is a polynomial modification with the same order of the teleparallel modification in Eq.~\eqref{themodel}. When $\alpha=0$, $\LL_2=\LL_1$.

Varing the action \eqref{themodel} with respect to $h^a_M$, we obtain the equation $L^{\,\,\,M}_{N}+k K^{\,\,\,M}_{N}=-{\Theta}^{\,\,\,M}_N$ with
\bes
\ben
L^{\,\,\,M}_{N}&=&h^{-1}\partial_Q (h S_N^{\,\,\,MQ})
- \tilde \Gamma^{Q}_{\,\,\,SN} S_Q^{\,\,\,MS}-\frac14 \delta_N^{\,\,\,M} T, \\
K^{\,\,\,M}_{N}&=&h^{-1}nT^{n-1}\partial_Q (h S_N^{\,\,\,MQ}) -nT^{n-1} \tilde \Gamma^{Q}_{\,\,\,SN} S_Q^{\,\,\,MS}
\nonumber \\
&&+n (n-1) T^{n-2}  S_N^{\,\,\,MQ}\partial_Q T
-\frac14 \delta_N^{\,\,\,M}T^n. 
\een
\ees
The energy-momentum tensor for a general scalar field is ${\Theta}_{MN}=\partial_M \phi \partial_N \phi {\cal L}_X-\eta_{ab}h^a_{\,\,\,M}h^{b}_{\,\,\,N}{\cal L}.$

Now, by varying the Eq.~\eqref{themodel} with respect to $\phi$, we get
\be\label{EqscalarPHI}
\frac{1}{h}\partial_M\left(h\,{\cal L}_X\partial^M\! \phi\right)+{\cal L}_\phi=0.
\ee
We study the case of a brane with the line element $ds^2=e^{2A} \eta_{\mu\nu} dx^\mu dx^\nu+dy^2$,  where $e^{2A}$ is the warp factor and $y=x_4$ is the extra dimension. For this, we choose $h^a_\mu=(e^{A},e^{A},e^{A},e^{A},1)$. As usual, in braneworld scenarios we suppose that $A$ and $\phi$ are static and only depend on $y$. Therefore we get $A=A(y)$ and $\phi=\phi(y)$. For this case, the equation of motion for the scalar field given by Eq.~\eqref{EqscalarPHI} becomes  
\be\label{phiEq1}
(\LL_X + 2\LL_{XX}X)\phi^{\prime\prime} - (2X\LL_{X\phi} - {\LL}_\phi) = -4 \LL_X \phi^\prime A^\prime.
\ee
Here, we take $X=-\phi^{\prime2}/2$ and the prime stands for the derivate with respect to $y$. The teleparallel gravity equations lead to the following equations
\bes\label{Einsteiss}
\ben
(1+C_nk A^{\prime(2n-2)})A^{\prime 2}&=&\frac13 (\LL - 2 \LL_X X), \label{Eeq1}\\
  (1+nC_n k A^{\prime(2n-2)}) A^{\prime\prime}&=& \frac43 \LL_X X, \label{Eeq2}
\een
\ees
where $C_n=(-1)^{n-1}2^{2n-2}3^{n-1} (2n-1)$ and the prime represents the derivative with respect to the argument. Some values for $C_n$ are $C_2=-36$, $C_3=720$, $C_4=-12096$. We take ${\cal L}_X\geq 0$. As $X$ and $A^{\prime\prime}$ are non-positive, we have the constraint $1+nC_n k A^{\prime(2n-2)}\geq 0$.  Sometimes this restricts the values on $k$. It is obvious that for $n=0$, the standard gravity equations~\cite{Brane4} are restored. 

The energy density is 
\be\label{rhodensity}
\rho(y)=-e^{2A(y)} {\cal L}.
\ee
By using Eqs.\eqref{Einsteiss}, we can write
\ben
\rho(y)&=&-\frac32\frac{d}{dy}\left[e^{2A}\left(A^\prime+C_n n k A^{\prime (2n-1)}\right)\right]\nonumber\\&&-\frac{3C_n (n-1)k}{2n-1}  e^{2A}A^{\prime(2n)},
\een 
with $n\neq 1/2$. For $k=0$, this expression is a total derivative, as seen in \cite{Brane2}. This result is not preserved for the model with torsion. If $k\neq 0$, the second term contributes to the energy. Thus, we can write
\be
{E}=-\frac{3C_n (n-1)k}{2n-1}  \int dy \,e^{2A}  A^{\prime(2n)}.
\ee
A similar result in the $F(R)$ braneworld scenario was obtained in Ref.~\cite{Bazeia:2013uva}.  

With the aim of introducing the first-order formalism, we choose the derivative of the warp factor with respect to the extra dimension to be a function of the scalar field:
\be
A^\prime=-\frac13 W(\phi).
\ee
Using this expression, the equation~\eqref{Eeq2} turns
\be\label{firstorderGeneral}
  \LL_X \phi^\prime=\frac{W_\phi}{2}\left(1+\frac{nC_n k}{3^{2n-2}} W^{2n-2}\right). 
\ee
This first order equation depends on the field $\phi$ and its derivative $\phi^\prime$. When possible, we will express $\phi^\prime=\phi^\prime(\phi)$. The equation \eqref{Eeq1} imposes a constraint to the Lagrangian density: 
\be\label{constraint}
(\LL - 2 \LL_X X)_{\phi^\prime=\phi^\prime(\phi)}=\left(1+\frac{C_nk}{3^{2n-2}} W^{2n-2}\right)\frac{W^2}{3}.
\ee
It is not difficult to show that these
first equations solve the second-order equations of motion \eqref{phiEq1} and \eqref{Eeq2}. 
We can express the equation \eqref{firstorderGeneral} as
\be\label{tildeW}
\!\!\LL_X \phi^\prime=\frac12 \tilde W_\phi, 
\ee
where
\be
\tilde W(\phi)=W(\phi) + \frac{nC_n k}{3^{2n-2}(2n-1)}(W(\phi))^{2n-1}.
\ee
Note that the Eq.~\eqref{tildeW} has the same form of the Eq.~\eqref{firstorderGeneral} with $n=0$. However, this simplification cannot be done in Eq.\eqref{constraint}. Therefore, it is not possible to reduce the first-order formalism to the equations of the standard gravity case~\cite{Brane4}. 

\section{Examples}

Let us now illustrate the general situation investigating some explicit models.
 
\subsection{Model I}

Firstly we investigate  the standard Lagrangian density $\LL_1$ given by equation \eqref{standardL}. In this case, from the first order equation \eqref{firstorderGeneral} we get
\bes
\be\label{firstorderStand}
   \phi^\prime=\frac{W_\phi}{2}\left(1+\frac{nC_n k}{3^{2n-2}} W^{2n-2}\right). 
\ee
The potential can be found by substituting this equation in the equation \eqref{constraint}
\ben\label{V_stand_XV}
V(\phi)&=&\frac{W_\phi^2}{8}\left(1+\frac{nC_n k}{3^{2n-2}} W^{2n-2}\right)^2 \nonumber\\&&\!-\frac{W^2}{3}\left(1+\frac{C_nk}{3^{2n-2}} W^{2n-2}\right).
\een
\ees
When $k=0$, as Eqs. \eqref{firstorderStand} and \eqref{V_stand_XV} becomes $\phi^\prime={W_\phi}/{2}$ and $V(\phi)={W_\phi^2}/{8}-{W^2}/{3}$, we have that the standard scenario is restored. The minima of the potential $v_i$ can be found solving the equation
\be
{W_\phi}=0 \,\,\,\,\,\,\, {\rm or} \,\,\,\,\,\,\,  W^{2n-2} =-\frac{3^{2n-2}}{nC_n k}.
\ee 
If we compare this equation with the Eq.~\eqref{firstorderStand}, we show that the solution has asymptotic behavior for these minima.   

\begin{figure}[t]
\includegraphics[width=4.2cm]{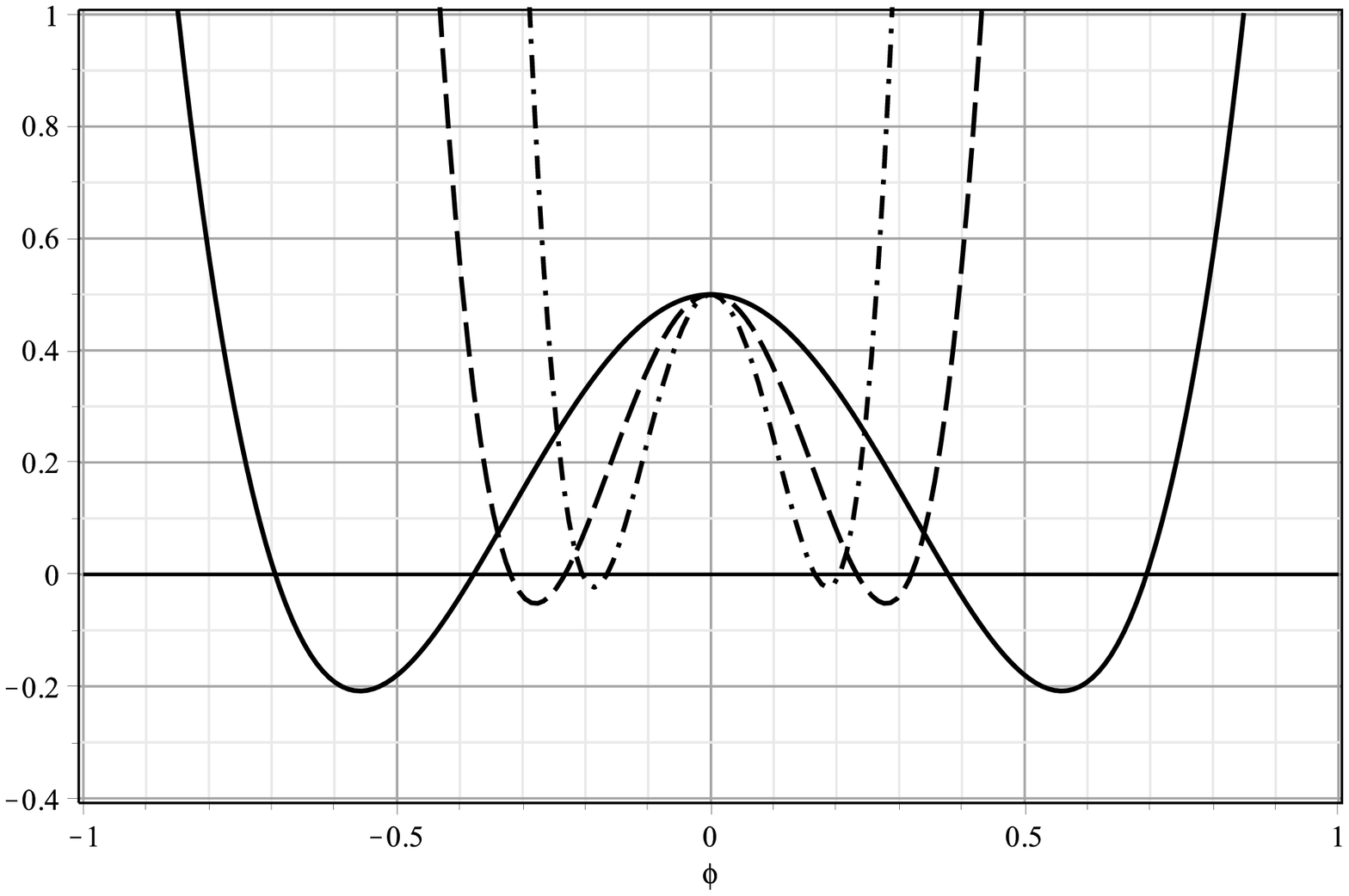}
\includegraphics[width=4.2cm]{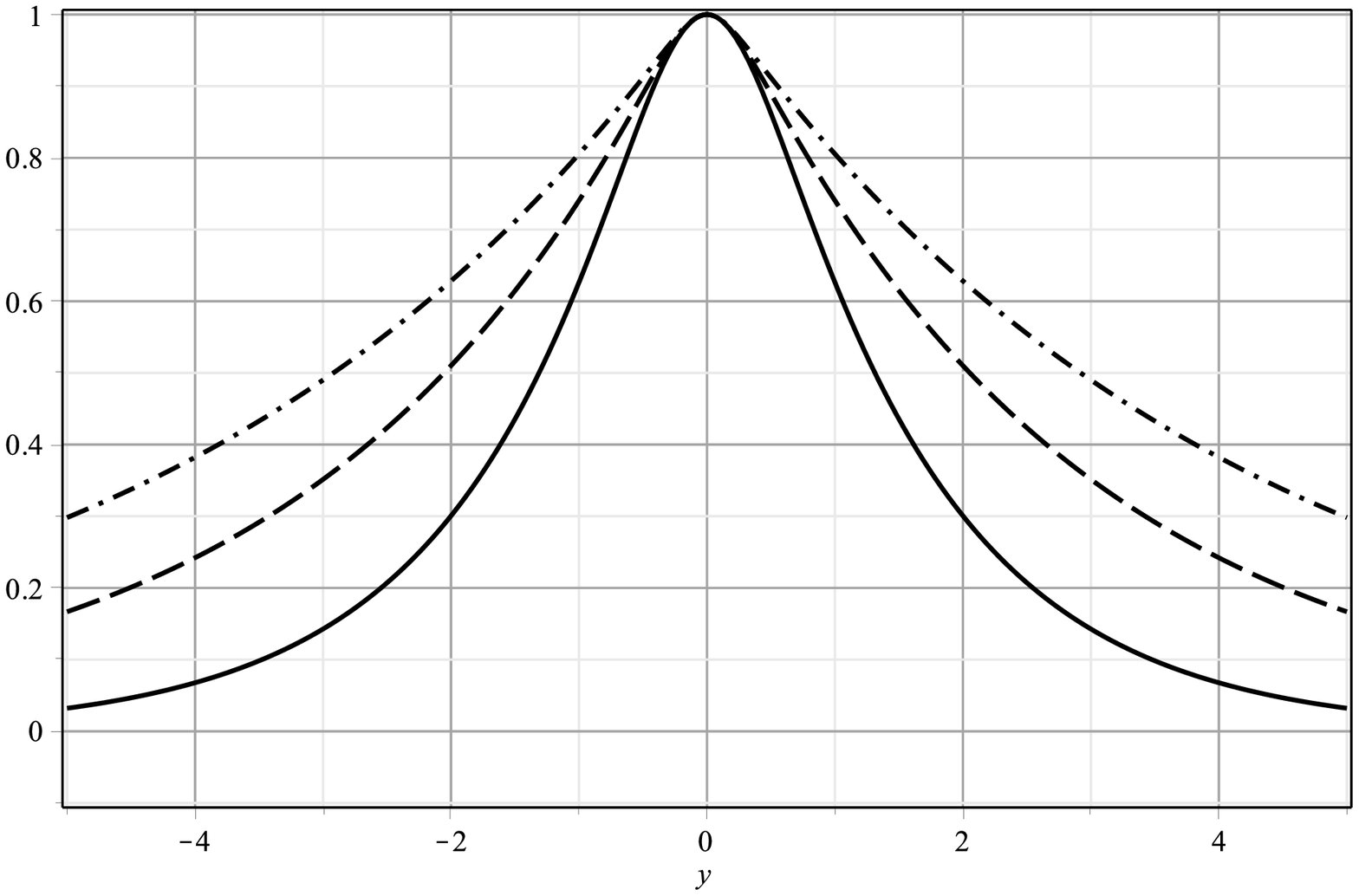}
\includegraphics[width=4.2cm]{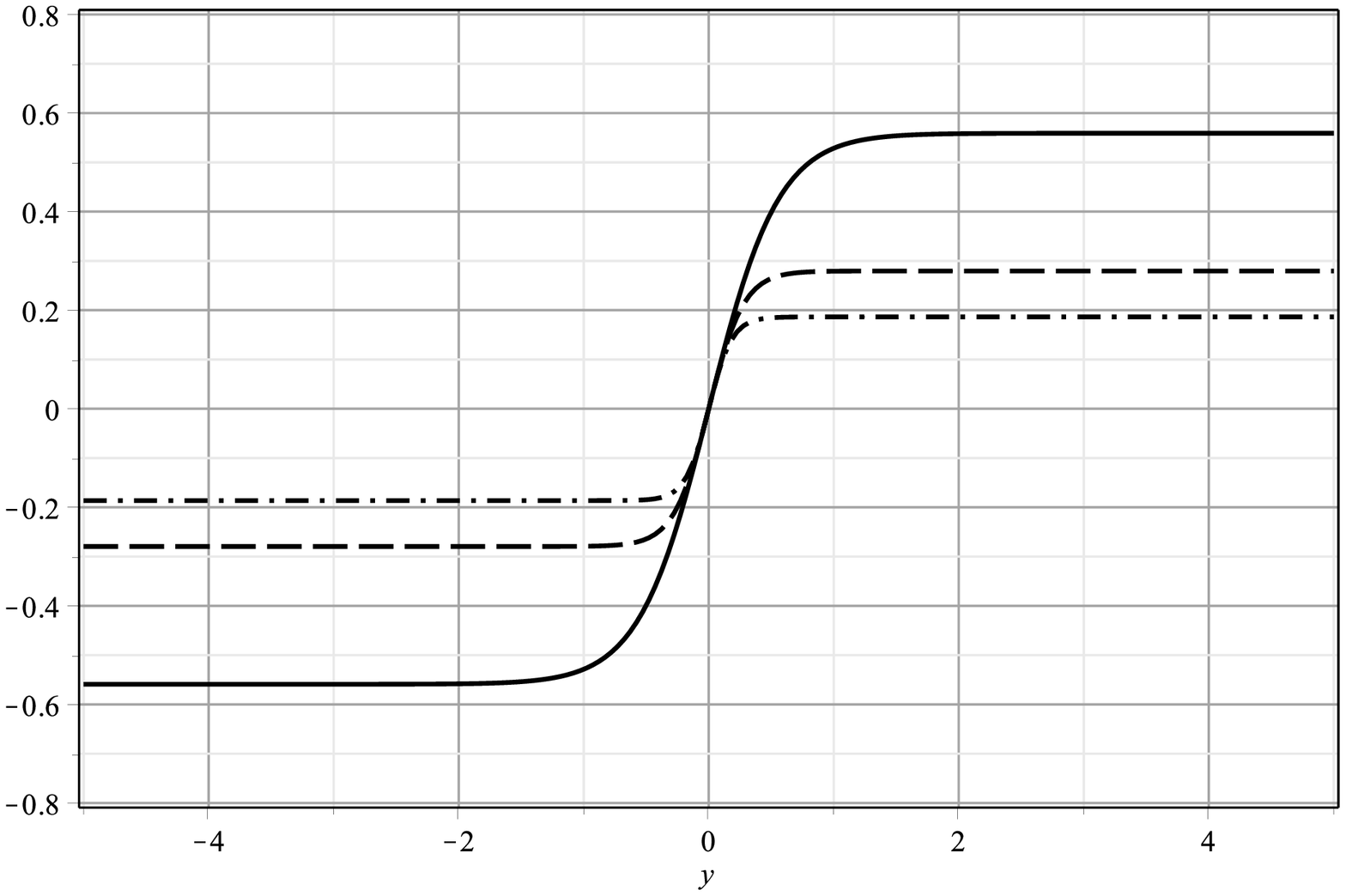}
\includegraphics[width=4.2cm]{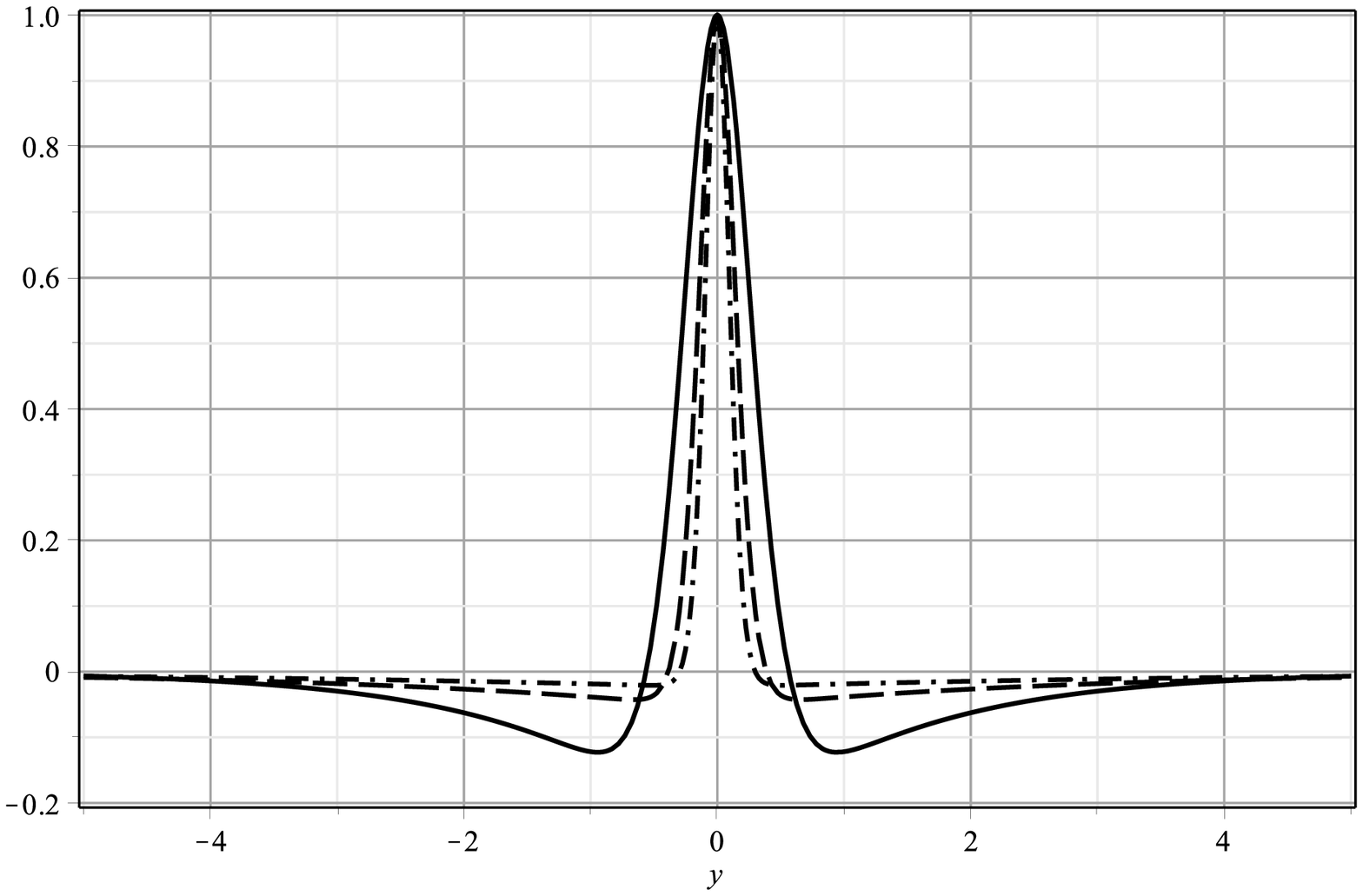}
\caption{Plots of potential \eqref{potential12}, warp factor \eqref{wcase1}, kink solution \eqref{case1} and energy density \eqref{rcase1}, respectively, in top left, top right, lower left and lower right panels. $n=2$,$k=0.1,k=0.4,k=0.9$.\label{FIG1}}    
\label{fig1}
\end{figure}

The energy density \eqref{rhodensity} is
\bes
\ben
\rho(y)&=&e^{2A(y)} \left(\frac12 \phi^{\prime2}+V(\phi)\right)\\
\nonumber &=&e^{2A(y)}\Bigg[\frac{W_\phi^2}{4}\left(1+\frac{nC_n k}{3^{2n-2}} W^{2n-2}\right)^2 \!\\&&-\frac{W^2}{3}\left(1+\frac{C_nk}{3^{2n-2}} W^{2n-2}\right)\Bigg].
\een
\ees

We take $W(\phi)$ as a linear function $W(\phi)=2\phi$. The potential and the first order equation \eqref{firstorderStand} can be written as
\ben\label{firstorderStandAA}
   \phi^\prime&=&1-B\phi^{2n-2}, \\
V(\phi)&=&\frac{1}{2}\left(1-B \phi^{2n-2} \right)^2\!\! -\!\frac{4\phi^2}{3}\!\left(\!1-\frac{B\phi^{2n-2}}{n} \right).
   \een
where $B=-{nC_n }{3^{2-2n}} 2^{2n-2}k$ has to be positive. Note that local maximum point is $\phi_0=0$, with $V(\phi_0)=1/2$, while the the minima are $\phi_\pm=\pm B^{\frac{1}{2-2n}}$, with $V(\phi_\pm)=-(4/3n)(n-1)\phi^{2}_{\pm}$. The solution is found in implicit form
\be
\phi \,\times\, 
{\mbox{$_2$F$_1$}(1,{(2n-2)}^{-1};\,1+{(2n-2)}^{-1};\,(B\phi)^{2n-2})}
=y
\ee
where $F$ is a hypergeometric function.

For  $n=2$, we take $B=32k$. The potential is written as
\be\label{potential12}
V(\phi)=\frac12 - \left(\!32k+\frac43\!\right)\!  \phi^2 + 64 k  \left(\! 8k +\frac13 \!\right) \!\phi^4,
\ee 
which is a sixth order polynomial. The first order equation is $\phi^\prime=1-32k \phi^2$. Only for $k>0$, this equation leads to a topological solution: 
\bes\label{case1}
\be
\phi(y)=\frac{1}{4\sqrt{2k}}\tanh(4\sqrt{2k}\,y).
\ee 
The ratio between the amplitude and thickness of the kink solution does not depends on $k$. The warp function is
\be\label{wcase1}
A(y)=\frac{1}{48k}\ln(\sech(4\sqrt{2k}\,y)).
\ee
\ees
In both solutions, we take $\phi(0)=A(0)=0$. These results were obtained in Ref.~\cite{Yang:2012hu}.  

The energy density is
\be\label{rcase1}
\rho(y)\!=\!S^{2p} \left[-p+\left(1+p\right)S^4\right],
\ee
where $S=\sech(4\sqrt{2k}\,y)$ and $p=(48k)^{-1}$. The energy density has $\rho(0)=1$ and $\rho^\prime(0)>0$ for any value of $k$. Thus, it does not split the brane. In the limit $k=0$, the kink solution is a straight line with $\phi^\prime=1$, the warp factor is constant and the energy density is not localized. Therefore, this solution only has significance for $k\neq 0$. In Fig.~\ref{FIG1}, we plot the profile of the potential $V(\phi)$, the warp factor $e^{2A(y)}$, the kink solution $\phi(y)$ and the energy density $\rho(y)$, for some values of $k$.   

\begin{figure}[t]
\includegraphics[width=4.25cm]{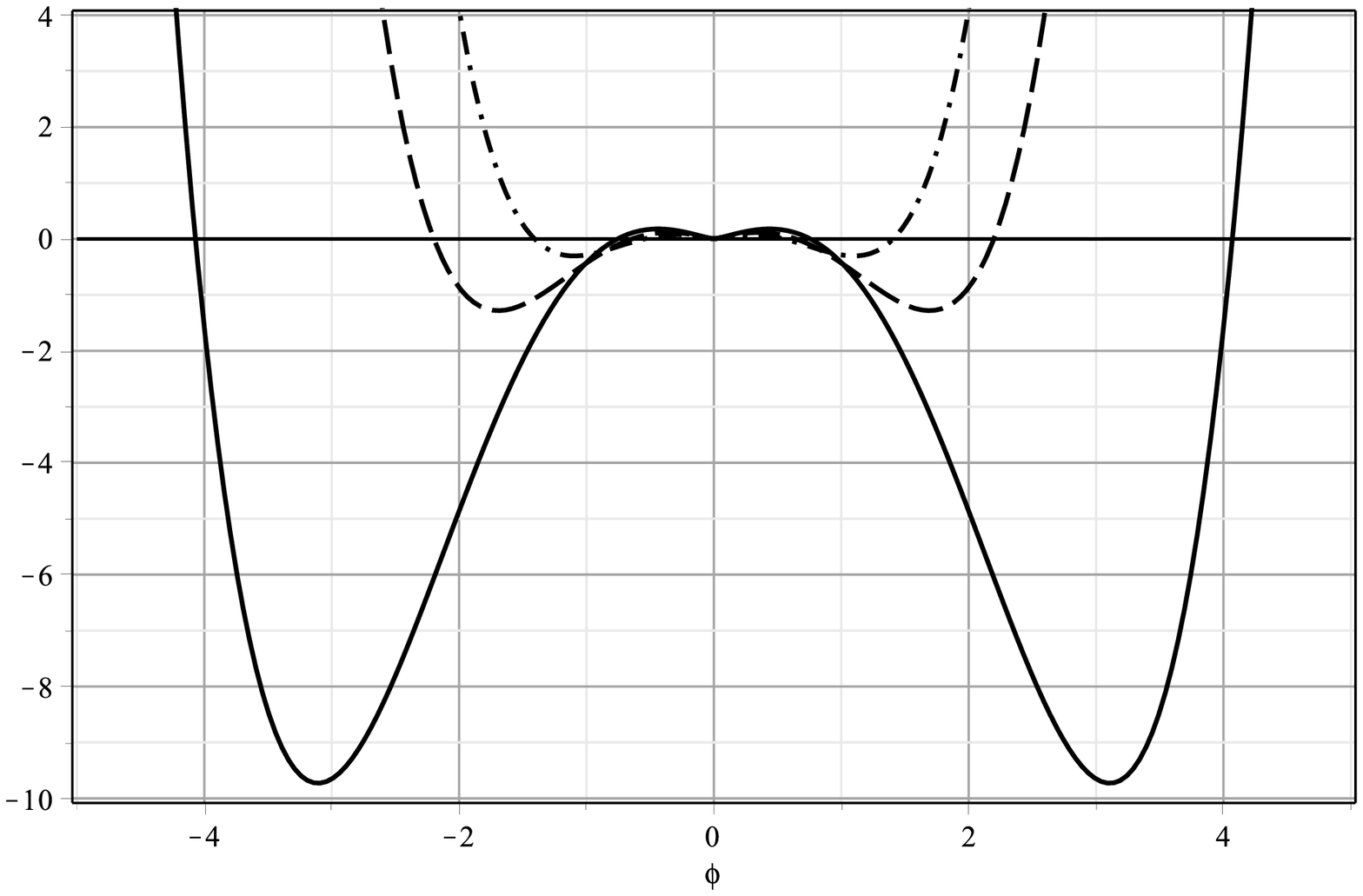}
\includegraphics[width=4.25cm]{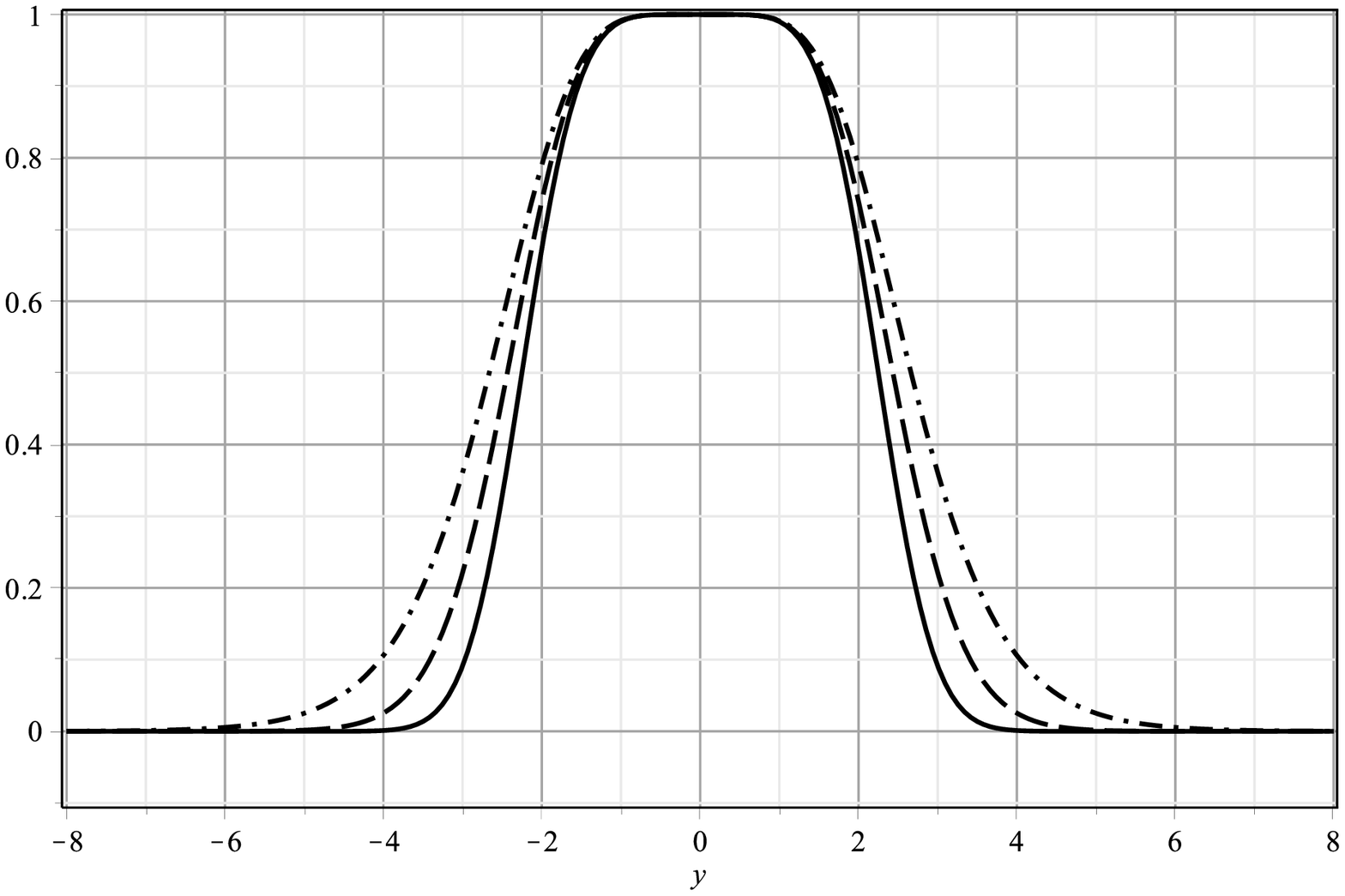}
\includegraphics[width=4.25cm]{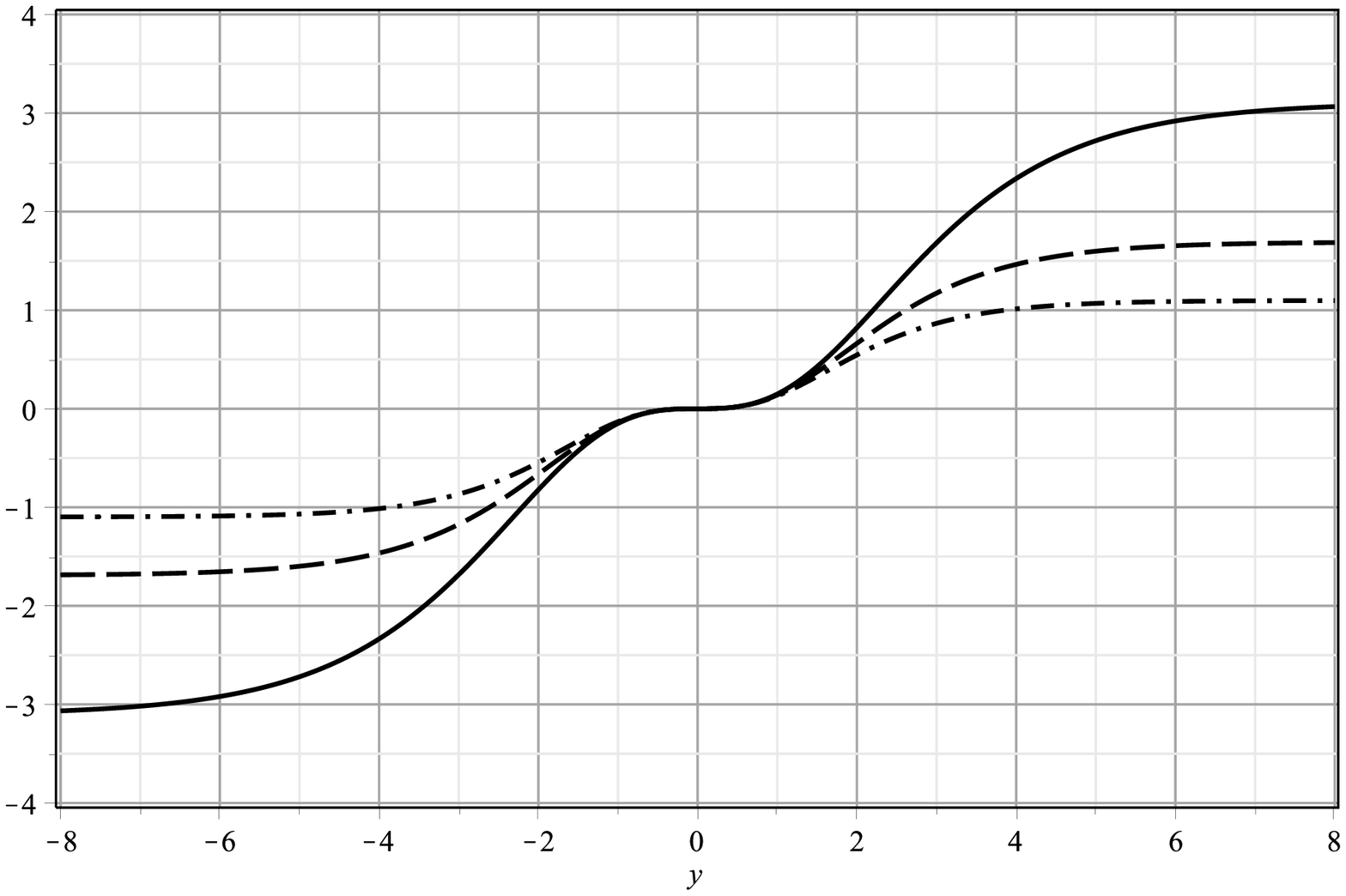}
\includegraphics[width=4.25cm]{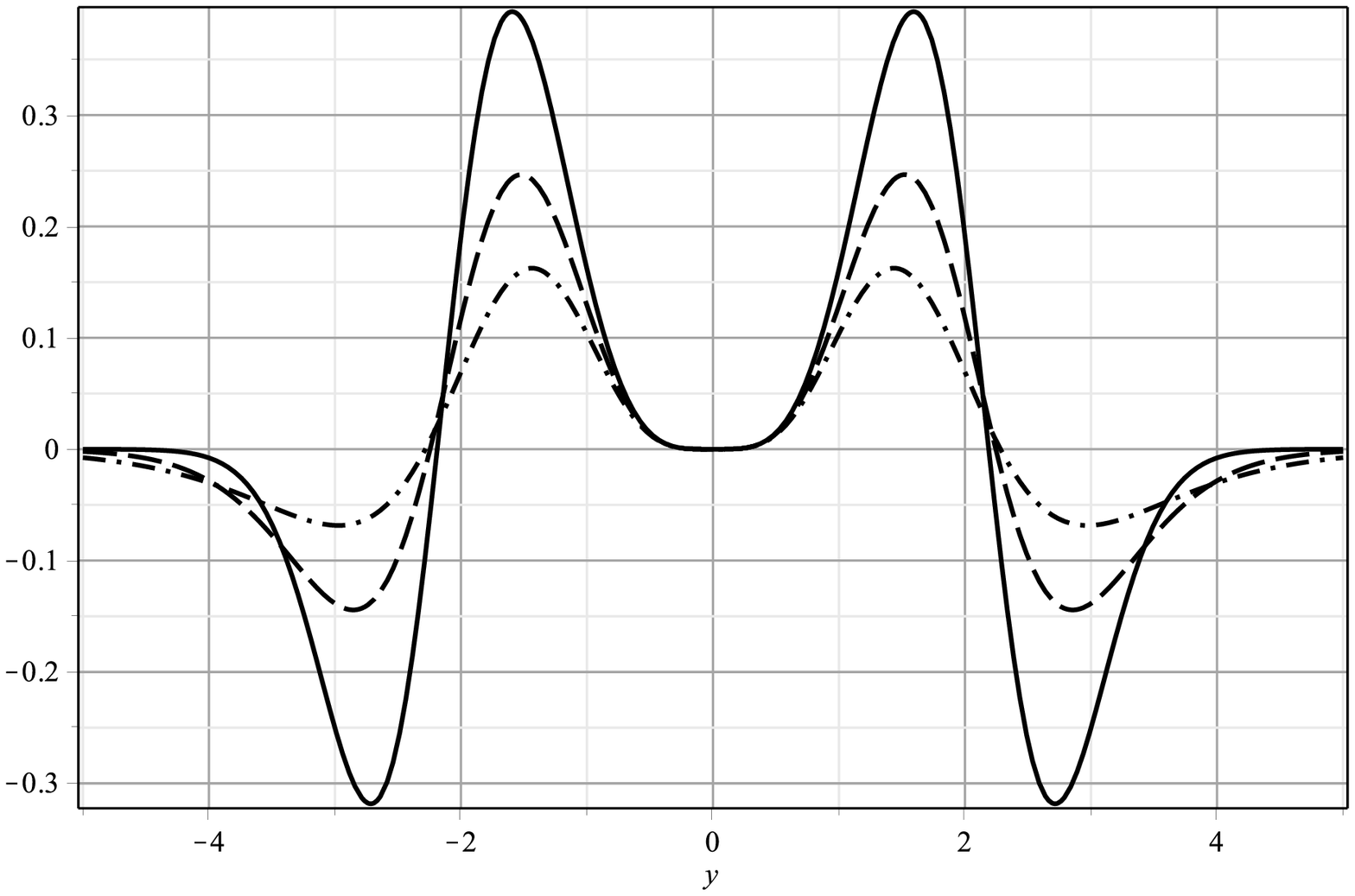}
\caption{Plots of potential \eqref{Vcase2}, warp factor \eqref{Acase2}, kink solution \eqref{Pcase2} and energy density \eqref{Rcase2}, respectively, in top left, top right, lower left and lower right panels. $n=6/5$, $k=0.2,k=0.3,k=0.4$.\label{FIG2}}
\label{fig1}
\end{figure}

Now we choose another function $W(\phi)=2\phi^{2/n}$, dependent on $n$, with $1< n \leq 2$. The potential and the first order equation are written as 
\bes
\ben
V(\phi)&=&\frac{2}{n^2} \phi^2 \left(\phi^{\frac{2-2n}{n}}\nonumber-B_n\phi^{\frac{2n-2}{n}}\right)^2\\&&-\frac{4}3 \left(\phi^{\frac4n}-\frac{B_n}{n}\phi^4\right), \label{Vcase2}\\
\phi^\prime &=& \frac{2}{n}\phi \left(\phi^{\frac{2-2n}{n}}-B_n \phi^{\frac{2n-2}{n}} \right),
\een
\ees
where $B_n=(-1)^{n}3^{1-n}\, 2^{4n-4}\,n (2n-1) k$ must be positive. For $n=2$, we obtain the previous case given by \eqref{case1}. For $n \neq 2$, the potential has a local minimum $\phi_0=0$, with $V(\phi_0)=0$ and two global minima  $\phi_\pm=\pm B_n^{-\frac{n}{4(n-1)}}$, with $V(\phi_\pm)=-4B_n^{-\frac{n}{n-1}}(n-1)/(3n)$.

The solution is
\ben\label{solutionN}
\phi(y)&=& \phi_+\tanh^{\frac{n}{2(n-1)}} \!\!\left( \,\frac{y}{\delta_n}\right),
\een
with $\delta_n^{-1}={ 4 (n-1)}\sqrt{B_n}/n^2$. In general, the warp factor is expressed as an integral
\ben
A(y)&=&-\frac{2\phi_+^{\frac1n}}3  \int dy\tanh^{\frac{1}{n-1}} \!\!\left(\,\frac{y}{\delta_n}\right).\een 
For $1<n<4/3$, the solution \eqref{solutionN} is a double-kink. For simplicity, we choose $n=6/5$. In this case, we can write the solutions as
\bes
\ben\label{Pcase2}
\phi(y) &=& \phi_+\tanh^3\left(\frac{y}{\delta_{6/5}}\right),\\
A(y) &=& \frac{6\phi_+^2}{5}\bigg[\ln(S)+\frac{3}{4}-S^2+\frac{1}{4}S^4\label{Acase2},
\bigg],
\een
\ees
where $S=\sech\!\left({y}/{\delta_{6/5}}\right)$. We take $\phi(0)=0$ and $A(0)=0$. Note that $\phi^\prime(0)=0$ and $A^\prime(0)=0$, which is the same behavior of the two-kink brane~\cite{Bazeia:2003aw}. In general, a two-kink solution describes branes engendering internal structure.  We can see this if we calculate the energy density, which is given by
\be\label{Rcase2}
\rho(y)\!=\!e^{2A(y)}\Big(\frac{25S^4(1-S^2)^2}{9B_{6/5}^2}
-\frac{2(5S^2+1)(1-S^2)^5}{9B^5_{6/5}}\Big).\ee
The brane has two symmetric maxima points, showing the splitting phenomenon. In the center of the brane, the energy density vanishes. In Fig.~\ref{FIG2}, we plot the profile of the potential $V(\phi)$, the warp factor $e^{2A(y)}$, the kink solution $\phi(y)$ and the energy density $\rho(y)$, for some values of $k$.   
All the solutions of this model exist only for nonzero $k$.


\subsection{Model II}

Now, we exemplify the first-order formalism for a nonstandard Lagrangian density for the scalar field. We take $\LL_2$ as it was given by Eq.\eqref{modifiedL}. In this case, using the first-order equation \eqref{firstorderGeneral}, we get  
\be\label{eqABCD}
\left[1+\alpha n b \left(1-\frac{b\phi^{\prime2}}2\right)^{n-1}\right]\phi^\prime=\frac{W_\phi}{2}\left(1+\frac{nC_n k}{3^{2n-2}} W^{2n-2}\right). 
\ee
In order to obtain $\phi^\prime$ as function of the $W(\phi)$ and $W_\phi$, we must solve an algebraic equation with degree $2n-1$. To avoid this, we impose that a possible solution for this equation is 
\be
\phi^\prime=\frac{W_\phi}2,
\ee 
which is the equation \eqref{eqABCD} when $k=\alpha=0$. By substituting this equation in the Eq. \eqref{eqABCD}, we obtain the following constraint for the $W(\phi)$ function
\be
\frac{1}{9}\left(\frac{C_n k}{\alpha b} \right)^{\frac{1}{n-1}}W^2+\frac{b}{8} W_\phi^2=1,
\ee
whose solution is 
\be
W(\phi)=3\left(\frac{\alpha b}{C_n k}\right)^{\frac{1}{2(n-1)}}\sin\left(\frac{2\sqrt{2}}{3\sqrt{b}} \left(\frac{C_n k}{\alpha b}\right)^{\frac{1}{2(n-1)}}\phi\right)
\ee
for $k \alpha C_n>0$. For simplicity, we choose $b=4/3$,$n=2$ and $\alpha=-27k$. With this, we can rewrite $W(\phi)=3\sin(\sqrt{2/3}\,\phi)$. The Eq. \eqref{constraint} allows us to write the potential as 
\ben
V(\phi)&=&{\frac {15}{4}}\, \cos^2 (\sqrt{2/3}\,\phi)  -3 \\&&- 27k \big(4
-10 \cos^2 (\sqrt{2/3}\,\phi) + 28\cos^4 (\sqrt{2/3}\,\phi) \big)\nonumber
\een
with $-1/36<k<1/792$. This potential have an infinite number of minima with $v_j=(\sqrt{3/2}\,\pi/2)j$, where $j$ is integer. The solution that connects the central minima is such that 
\bes
\ben
\phi(y)&=&\sqrt{2/3}\arcsin(\tanh(y))\\
A(y)&=&\ln(S)\\
\rho(y)&=&{S^2}[4S^2-3+54k(2S^4+3S^2-2)]
\een
\ees
where $S=\sech(y)$. Another way to investigate the equation is to take $\alpha$ and $k$ very small and to discard second order contributions of these parameters to obtain the solutions for a generic $W(\phi)$. 

 \section{Final Remarks} 
 
In this work, we introduced the first-order formalism to find analytical solutions for thick branes in the modified teleparallel gravity coupled to the scalar field, with the Lagrangian density in the standard and generalized forms, given by ${\LL_1}$ and $\LL_2$, respectively. This study is a natural continuation of the program to investigate the scalar field behavior under the presence of generalized dynamics with the gravity introduced in Ref. \cite{Brane4}. We found analytical profiles with brane splitting for any value of the parameter $k$. These solutions are two-kink structures and, in the brane context, the profile engenders vanishing derivative at its center, with the energy density also vanishing there. 

The same procedure can be used to investigate other cases involving thick brane solution in generalized gravity. For instance, we can take general $F(T)$ functions, besides the polynomial form. An interesting way is to choose small modifications in usual gravity, $F(R,T)=R+\alpha(F_1(R)+F_2(T))$, where $\alpha$ is a small parameter and $F_1(R)$ and $F_2(T)$ are generic functions of $R$ and $T$. This could extend the study done in Ref. \cite{Bazeia:2013uva} to other scenarios of current interest.
 
{\it Acknowledgments.} The author  like to thank D. Bazeia, L. Losano and  J.G. Ramos, for useful comments; and CAPES and CNPq, for partial financial support.

\end{document}